# Data Grid Concepts for Data Security in Distributed Computing

A.S.Syed Navaz,   C.Prabhadevi,   V.Sangeetha

Department of Computer Applicatons,
Muthayammal College Of Arts & Science, Namakkal, India.

## ABSTRACT
Data grid is a distributed computing architecture that integrates a large number of data and computing resources into a single virtual data management system. It enables the sharing and coordinated use of data from various resources and provides various services to fit the needs of high-performance distributed and data-intensive computing. Here data partitioning and dynamic replication in data grid are considered. In which security and access performance of a system are efficient. There are several important requirements for data grids, including information survivability, security, and access performance. More specifically, the investigation is the problem of optimal allocation of sensitive data objects that are partitioned by using secret sharing scheme or erasure coding scheme and replicated. DATA PARTITIONING is known as the single data can be divided into multiple objects. REPLICATION is known as process of sharing information. (i.e.) storing same data in multiple systems. Replication techniques are frequently used to improve data availability. Single point failure does not affect this system. Where the data will be secured

## 1. INTRODUCTION
Data grid is a distributed computing architecture that integrates a large number of data and computing resources into a single virtual data management system. It enables the sharing and coordinated use of data from various resources and provides various services to fit the needs of high-performance distributed and data-intensive computing. Many data grid applications are being developed or proposed, such as DOD's Global Information Grid (GIG) for both business and military domains, NASA's Information Power Grid GMESS Health-Grid for medical services, data grids for Federal Disaster Relief, etc. These data grid applications are designed to support global collaborations that may involve large amount of information, intensive computation, real time, or non real time communication. Success of these projects can help to achieve significant advances in business, medical treatment, disaster relief, research, and military and can result in dramatic benefits to the society.

There are several important requirements for data grids, including information survivability, security, and access performance. For example, consider a first responder team responding to a fire in a building with explosive chemicals. The data grid that hosts building safety information, such as the building layout and locations of dangerous chemicals and hazard containment devices, can help draw relatively safe and effective rescue plans. Delayed accesses to these data can endanger the responders as well as increase the risk to the victims or cause severe damages to the property. At the same time, the information such as location of hazardous chemicals is highly sensitive and, if falls in the hands of terrorists, could cause severe consequences. Thus, confidentiality of the critical information should be carefully protected. The above example indicates the importance of data grids and their availability, reliability, accuracy, and responsiveness. Replication is frequently used to achieve access efficiency, availability, and information survivability. The underlying infrastructure for data grids can generally be classified into two types cluster based and peer-to-peer Systems.

In pure peer-to-peer storage systems, there is no dedicated node for grid applications (in some systems, some servers are dedicated). Replication can bring data objects to the peers that are close to the accessing clients and, hence, improve access efficiency. Having multiple replicas directly implies higher information survivability. In cluster-based systems, dedicated servers are clustered together to offer storage and services. However, the number of clusters is generally limited and, thus, they may be far from most clients. To improve both access performance and availability, it is necessary to replicate data and place them close to the clients, such as peer-to-peer data caching. As can be seen, replication is an effective technique for all types of data grids. Existing research works on replication in data grids investigate replica access protocols resource management and discovery techniques replica location and discovery algorithms and replica placement issues.

Replication of keys can increase its access efficiency as well as avoiding the single-point failure problem and reducing the risk of denial of service attacks, but would increase the risk of having some compromised key servers. If one of the key servers is compromised, all the critical data are essentially compromised. Beside key management issues, information leakage is another problem with the replica encryption approach. Generally, a key is used to access many data objects. When a client leaves the system or its privilege for some accesses is revoked, those data objects have to be re encrypted using a new key and the new key has to be distributed to other clients. If one of the data storage servers is compromised, the storage server could retain a copy of the data encrypted using the old key. Thus, the content of long-lived data may leak over time. Therefore, additional security mechanisms are needed for sensitive data protection. In this paper, we consider combining data partitioning and replication to support secure, survivable, and high performance storage systems.

## 2. EXISTING SYSTEM

The intrusion tolerance concept and data partitioning techniques can be used to achieve data survivability as well as security. The most commonly used schemes for data partitioning include secret sharing and erasure coding. Both schemes partition data into shares and distribute them to different processors to achieve availability and integrity. Secret sharing schemes assure confidentiality even if some shares (less than a threshold) are compromised. In erasure coding, data shares can be encrypted and the encryption key can be secret shared and distributed with the data shares to assure confidentiality. However, changing the number of shares in a data partitioning scheme is generally costly. When it is necessary to add additional shares close to a group of clients to reduce the communication cost and access latency, it is easier to add share replicas. Thus, it is most effective to combine the data partitioning and replication techniques for high-performance secure storage design.

## 3. PROPOSED SYSTEM

We consider data partitioning (both secret sharing and erasure coding) and dynamic replication in data grids, in which security and data access performance are critical issues. More specifically, we investigate the problem of optimal allocation of sensitive data objects that are partitioned by using secret sharing scheme or erasure coding scheme and/or replicated.
DATA grid is a distributed computing architecture that integrates a large number of data and computing resources into a single virtual data management system. It enables the sharing and coordinated use of data from various resources and provides various services to fit the needs of high-performance distributed and data-intensive computing.
Replication techniques are frequently used to improve data availability and reduce client response time and communication cost. One major advantage of replication is performance improvement, which is achieved by moving data objects close to clients. In full replication all servers keep a complete set of the data objects.

## 4. ARCHITECTURE

The grid Architecture involves these following concepts, Data grid, middleware, data fragmentation, data replication.

### 4.1. What is Data Grid?

A Data Grid is an architecture or set of services that enable individuals or groups of users the ability to access, modify and transfer extremely large amounts of geographically distributed data for research purposes. Data grids make this possible through a host of middleware applications and services that pull together data and resources from multiple administrative domains and then present it to users upon request. The data in a data grid can be located at a single site or multiple sites where each site can be its own administrative domain governed by a set of security restrictions as to who may access the data. Likewise, multiple replicas of the data may be distributed throughout the grid outside their original administrative domain and the security restrictions placed on the original data for who may access it must be equally applied to the replicas. Specifically developed data grid middleware is what handles the integration between users and the data they request by controlling access while making it available as efficiently as possible.

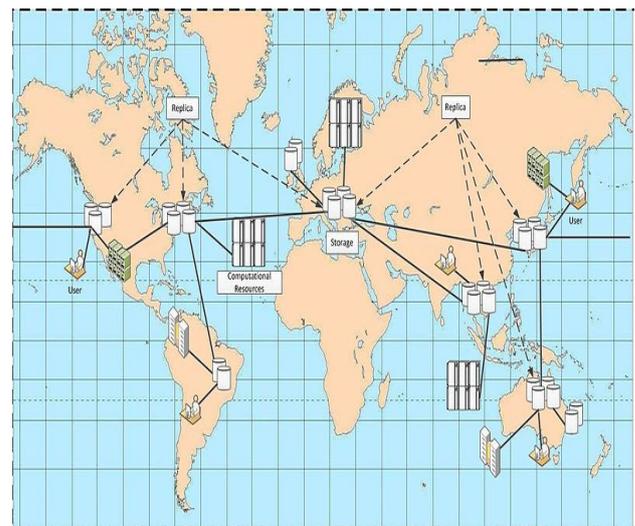

**Fig 1.Datagrid topology**

### 4.2. Middleware

Middleware provides all the services and applications necessary for efficient management of datasets and files within the data grid while providing users quick access to the datasets and files Data access services work hand in hand with the data transfer service to provide security, access controls and management of any data transfers within the data grid. Security services provide mechanisms for authentication of users to ensure they are properly identified. Common forms of security for authentication can include the use of passwords. Authorization services are the mechanisms that control what the user is able to access after being identified through authentication.

## 4.3. Data fragmentation

Data fragmentation is a process of division or mapping database where the database is broken down into number of parts then stored in the site or units of different computers in a data network, allowing for decision-making to data that has been divided. Data that has broken down is still possible to be combined again with the intention to complete the data collection. When doing fragmentation, data must meet several conditions for the fragment is correct, below is fragmentation principle.

*4.3.1. Completeness:* a unit of data that is still in the main part of the relationship, then the data must be in one fragment. When there is a relation, the distribution of the data must be an integral part of the relationship.

*4.3.2. Reconstruction*: an original relation can be reused or combined return of a fragment. When it has broken down, data is still possible to be combined again with no change in the structure of data.

*4.3.3. Disjointness:* data within the fragment should not be included in the other fragments in order to avoid redundancy of data, except for primary key attributes of vertical fragmentation.

## 4.4. Data replication

Replication is known as process of sharing information. (i.e.) storing same data in multiple systems.

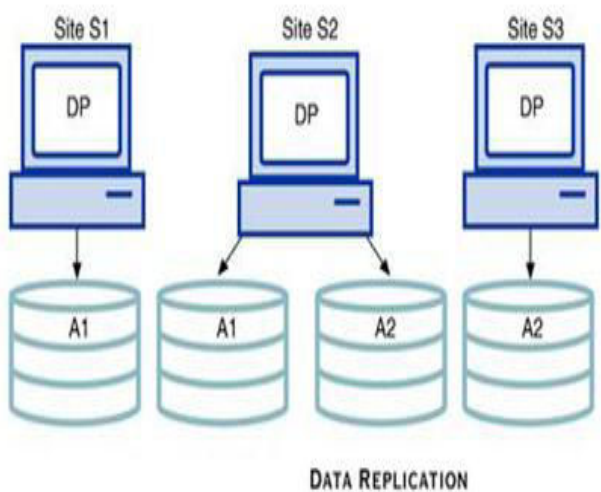

**Fig 2.Data Replication**

Suppose database A is divided into two fragments, A1 and A2. Within a replicated distributed database, the scenario depicted in the following Figure is possible: fragment A1 is stored at sites S1 and S2, while fragment A2 is stored at sites S2 and S3.

Replication techniques are frequently used to improve data availability reduce client response time and communication cost. Single point failure does not affect this system. Where the data will be secured. One major advantage of replication is performance improvement, which is achieved by moving data objects close to clients. In full replication all servers keep a complete set of the data objects.

Three replication scenarios exist: a database can be fully replicated, partially replicated, or un replicated.

*4.4.1. A fully replicated database:* stores multiple copies of each database fragment at multiple sites. In this case, all database fragments are replicated. A fully replicated database can be impractical due to the amount of overhead it imposes on the system

*4.4.2. A partially replicated database:* stores multiple copies of *some* database fragments at multiple sites. Most DDBMSs are able to handle the partially replicated database well.

*4.4.3. An unreplicated database:* stores each database fragment at a single site. Therefore, there are no duplicate database fragments.
In this project we have used a fully replicated database and a partially replicated database concepts.
Several factors influence the decision to use data replication:

*4.4.4. Database size:* The amount of data replicated will have an impact on the storage requirements and also on the data transmission costs. Replicating large amounts of data requires a window of time and higher network bandwidth that could affect other applications.

*4.4.5. Usage frequency:* The frequency of data usage determines how frequently the data needs to be updated. Frequently used data needs to be updated more often, for example, than large data sets that are used only every quarter.

*4.4.6. Costs:* including those for performance, software overhead, and management associated with synchronizing transactions and their components vs. fault-tolerance benefits that are associated with replicated data.

## 5. TASKS INVOLVED IN THIS PROJECT

There are four modules involved in this project
- Network Module
- Dynamic randomization process
- Secure data share
- Replication data grids

## 5.1. Network Module

Client-server computing or networking is a distributed application architecture that partitions tasks or workloads between service providers (servers) and service requesters, called clients. Often clients and servers operate over a computer network on separate hardware. A server machine is a high-performance host that is running one or more server programs which share its resources with clients. A client also shares any of its resources; Clients therefore initiate communication sessions with servers which await (listen to) incoming requests.

## 5.2. Dynamic Randomization Process

The delivery of a packet with the destination at a node in order to minimize the probability that packets are eavesdropped over a specific link, a randomization process for packet deliveries, in this process, the previous next-hop for the source node s is identified in the first step of the process. Then, the process randomly picks up a neighboring node as the next hop for the current packet transmission. The exclusion for the next hop selection avoids transmitting two consecutive packets in the same link, and the randomized pickup prevents attackers from easily predicting routing paths for the coming transmitted packets.

## 5.3. Secure data share

Secure data partitioning (both secret sharing and erasure coding) and dynamic replication in data grids, in which security and data access performance are critical issues. More specifically, we investigate the problem of optimal allocation of sensitive data objects that are partitioned by using secret sharing scheme or erasure coding scheme and/or replicated. We consider achieving secure, survivable, and high-performance data storage in data grids. To facilitate scalability, we model the peer-to-peer data grid as a topology. Our goal is to replicate the data shares and allocate them to different nodes in the data grid to minimizing cost. We decompose the allocation problem into two sub problems—intra cluster and inter cluster share allocation problems—and deal with them separately and independently

## 5.4 Replication Data Grids

Data grid is a distributed computing architecture that integrates a large number of data and computing resources into a single virtual data management system. It enables the sharing and coordinated use of data from various resources and provides various services to fit the needs of high-performance distributed and data-intensive computing.

Dynamic replication to achieve data survivability, security, and access performance in data grids. The replicas of the partitioned data need to be properly allocated to achieve the actual performance gains. We have developed algorithms to allocate correlated data shares in large-scale peer-to-peer data grids Replication is a natural solution for reducing the communication cost (as we have discussed) as well as sharing the access load. In peer-to-peer data grids, replica can be placed on widely distributed nodes to achieve better access performance and load sharing. In cluster-based data grids, caching data on widely distributed nodes is necessary (in addition to replication on cluster nodes) to achieve improved access performance and load sharing. Data partitioning can contribute to reduced storage cost. It has been shown that erasure coding-based schemes can greatly reduce the overall storage cost and effectively share the storage consumption.

## 6. SYSTEM DESIGN
### 6.1. System Architecture

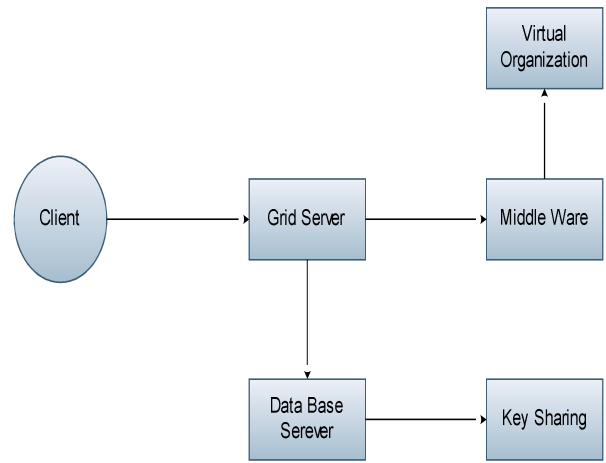

**Fig 3. System Architecture**

The client can secure the data through the grid server and it will be stored in data grid by data fragmentation and data replication concept, and also can access the data from grid server.

### 6.2. Use case Diagram

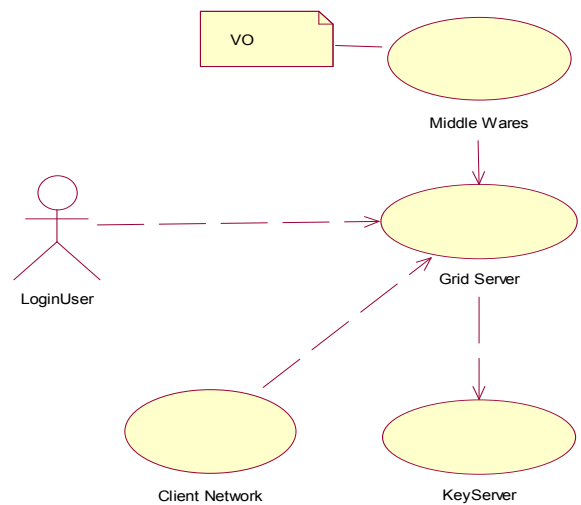

**Fig 4. Use case Diagram**

## 6.3. Data flow diagram

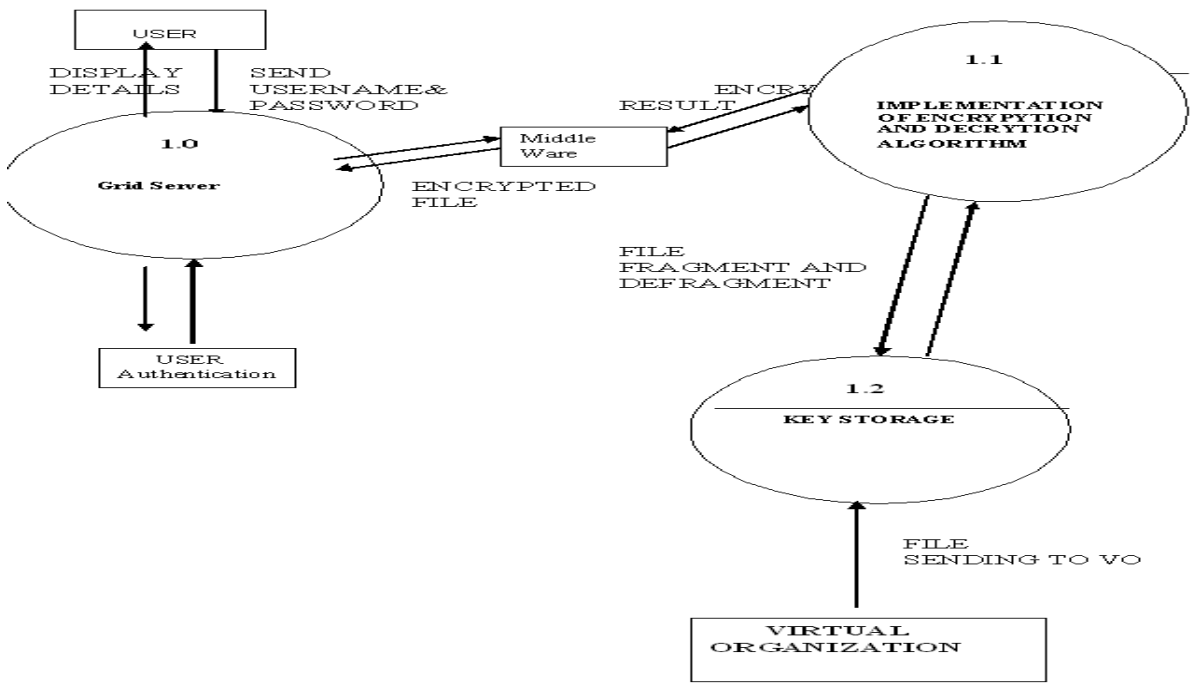

**Fig 5. Data Flow Diagram**

## 6.4. Sequence Diagram

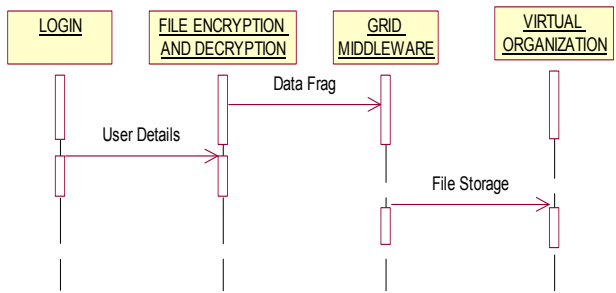

**Fig 6. Sequence diagram**

## 6.5. Collaborative Diagram

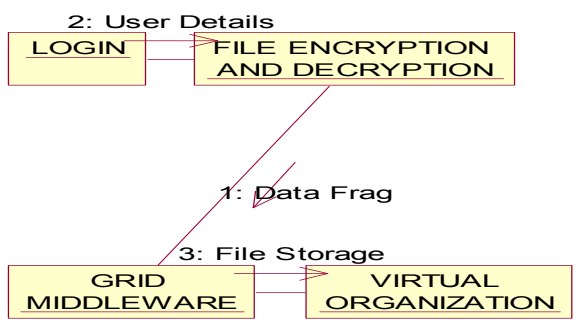

**Fig 7. Collaborative Diagram**

## 7. ADVANTAGES

- Data will be secured
- It enables the sharing and coordinated use of data from various resources and provides various services to fit the needs of high-performance distributed and data-intensive computing
- Replication techniques are frequently used to improve data availability and reduce client response time and communication
- Single point failure does not affect this system

## 8. CONCLUSION

We have combined data partitioning schemes (secret sharing scheme or erasure coding scheme) with dynamic replication to achieve data survivability, security, and access performance in data grids. The replicas of the partitioned data need to be properly allocated to achieve the actual performance gains. We have developed algorithms to allocate correlated data shares in large-scale peer-to-peer data grids. Data grid is a distributed computing architecture that integrates a large number of data and computing resources into a single virtual data management system. It enables the sharing and coordinated use of data from various resources and provides various services to fit the needs of high-performance distributed and data-intensive computing. Moreover, it may be desirable to consider multiple factors for the allocation of secret shares and their replicas. Replicating data shares improves access performance but degrades security. Having more share replicas may increase the chance of shares being compromised. Thus, it is desirable to determine the placement solutions based on multiple objectives, including performance, availability, and security.

## 9. FUTURE ENHANCEMENT

Now we applied only in Data Grid security. In future we can apply at any sort of business application to produce absolute development and with security enhancement.

## 11. AUTHOR'S PROFILE

**A.S.Syed Navaz** received BBA from Annamalai University, Chidambaram 2006, M.Sc Information Technology from KSR College of Technology, Anna University Coimbatore 2009, M.Phil in Computer Science from Prist University, Thanjavur 2010 and M.C.A from Periyar University, Salem 2010 .Currently he is working as an Asst.Professor in Department of BCA, Muthayammal College of Arts & Science, Namakkal. His area of interests are Computer Networks and Mobile Communications.

**C.Prabhadevi** received BCA from Sengunthar Arts & Science College, Periyar University 2005 , M.Sc Computer Science from Sengunthar Arts & Science College, Periyar University 2007, M.C.A from Periyar University, Salem 2009 and M.Phil in Computer Science from Prist University, Thanjavur 2012. Currently she is working as an Asst.Professor in Department of BCA, Muthayammal College of Arts & Science, Namakkal. Her area of interests are Digital Image Processing and Mobile Communications.

**V.Sangeetha** received B.Sc Computer Science from Mahendra Arts & Science College, Periyar University 2003, M.Sc Computer Science from Mahendra Arts & Science College, Periyar University 2005 and M.Phil in Computer Science from Periyar University, Salem 2008. Currently she is working as an Asst.Professor in Department of BCA, Muthayammal College of Arts & Science, Namakkal. Her area of interests are Computer Networks and Mobile Communications.